\documentclass[10pt,letterpaper]{article}
\usepackage{opex3}
\usepackage{cite}
\usepackage{graphicx}
\usepackage[hidelinks=true]{hyperref}
\usepackage{nicefrac}
\usepackage{ae} 

\begin{document}

\title{Influence of signal bandwidth on mode instability threshold of fiber amplifiers}

\author{Jesse J. Smith$^*$ and Arlee V. Smith}

\address{AS-Photonics, LLC, 6916 Montgomery Blvd. NE, Suite B8, Albuquerque, NM 87109 USA}

\email{$^*$jesse.smith@as-photonics.com}

\begin{abstract}
We show how signal linewidth affects the gain of stimulated thermal Rayleigh scattering (STRS) which is responsible for mode instability in fiber amplifiers. The gain is reduced if the coherence time of the signal is less than the group velocity induced walk off between modes LP$_{01}$ and LP$_{11}$. We derive equations for short pulses, linear chirps, and general periodic cases.
\end{abstract}

\ocis{(060.2320) Fiber optics amplifiers and oscillators; (060.4370) Nonlinear optics, fibers; (140.6810) Thermal effects; (190.2640) Stimulated scattering, modulation, etc}

\section{Introduction}

Yb-doped fiber amplifiers offer a promising route to multikilowatt operation with good beam quality in the near infrared. However, they are not without problems. Namely, nonlinear processes such as stimulated Brillouin scattering (SBS), stimulated Raman scattering (SRS) and stimulated thermal Rayleigh scattering (STRS) can compromise high power performance.

In fiber amplifiers, stimulated Brillouin scattering is an exponential gain process in which the forward-propagating signal pumps a backward-propagating Stokes wave. Because the signal and Stokes waves have different propagation velocities, phase modulating the signal wave reduces the SBS gain and raises the SBS threshold. Stimulated thermal Rayleigh scattering is also an exponential gain process in which a forward propagating signal in mode LP$_{01}$ pumps a forward-propagating Stokes wave in mode LP$_{11}$. These two waves also have different group velocities, so phase modulating the signal wave can again reduce the STRS gain and raise the mode instability threshold. In this paper, we will provide a physical explanation of the gain reduction and derive equations that allow qualitative calculation of the increase in threshold.

Currently, high power fiber amplifiers are often limited by transverse mode instability where power that is initially in the lowest order mode is suddenly transferred to higher order modes above a sharp output power threshold. This destroys beam quality. The physical process involves optical interference between the two modes, creating an irradiance grating. Pump light is absorbed more strongly in zones with higher irradiance, creating a heating pattern that mimics the irradiance grating. This leads to a temperature grating, and through the thermo-optic effect a refractive index grating which can couple light between the modes. One other requirement is a phase shift between the irradiance and refractive index grating, which is caused by a frequency-shift between modes. This is described in more detail in refs.\cite{Smith2011,Smith2012,Smith2013a,Smith2013e,Smith2013f,Hansen2012,Hansen2013,Hansen2014,Dong2013,Smith2014}.

Typical mode instability threshold powers are several hundred watts to a few kilowatts\cite{Eidam2011,Otto2012,Karow2012,Ward2012,Brar2014}. Previous work has explained how to maximize the threshold by minimizing seeding of the higher order mode by either eliminating amplitude modulation of the pump and signal \cite{Smith2012,Hansen2013}, or by minimizing spontaneous thermal Rayleigh scattering \cite{Smith2013e}. Larger increases in threshold can achieved by reducing STRS gain through gain saturation \cite{Smith2013f,Hansen2014}. Increasing the signal linewidth also reduces the STRS gain, provided the linewidth is sufficiently large. In this paper, we explore the influence of linewidth on STRS gain for several scenarios: short (picosecond) pulses, frequency-swept amplifiers, and a general periodic modulation.

\section{Short pulse amplifiers}\label{sec:pulses}

Mode instability has been observed in picosecond pulsed amplifiers with repetition rates of MHz\cite{Eidam2011}. The width of the picosecond pulses defines the linewidth of interest. The high repetition rate means the temperature profile, which takes milliseconds to develop, averages over many pulses. The Stokes shift due to STRS, which is on the order of one kilohertz, is also much smaller than the repetition rate.

In order for STRS coupling to occur, optical interference of the light in modes LP$_{01}$ and LP$_{11}$ is necessary. Pulses must exist in both modes which overlap in time for interference to occur. However, due to the different group velocities in the two modes, short pulses will cease to overlap after some propagation distance. We use this to set the criterion that for STRS suppression of short pulses to occur, the temporal walkoff between modes must be equal to or greater than the full-width half-maximum (FWHM) pulse duration. The critical limit on the pulse duration is thus equal to the temporal walkoff between modes. If the pulse duration is shorter than this critical value STRS gain is significantly diminished. The critical value for pulse duration implies a critical value for linewidth.

The group velocity walkoff is the difference in the times of flight from $z = 0$ to $z = L$, where $L$ is the length of the fiber. The time of flight for mode $j$ is 
\begin{equation}
T_j = \frac{L}{{\rm v}_{gj}},
\end{equation}
where ${\rm v}_{gj}$ is the group velocity of mode $j$.
The difference in times of flight over the full length of the fiber is
\begin{equation}
\Delta T = L \left[ \frac{1}{{\rm v}_{g1}} - \frac{1}{{\rm v}_{g2}} \right],
\end{equation}
so our criterion implies
\begin{equation}
\tau_{\rm c} = L \left[ \frac{1}{{\rm v}_{g1}} - \frac{1}{{\rm v}_{g2}} \right].
\end{equation}

Assuming the pulse is unchirped, the time-bandwidth product for Gaussian temporal profiles is
\begin{equation}
\label{eqn:time-bandwidth_gaussian} \tau \Delta \nu = 0.44,
\end{equation}
where $\tau$ is the temporal width (FWHM) of the pulse and $\Delta \nu$ is its bandwidth (FWHM). Then the critical bandwidth at which gain begins to be suppressed is given by 
\begin{equation}
\label{eqn:delta_nu_pulses}\Delta \nu_{\rm c} = \frac{0.44}{\tau_{c}} = \frac{0.44}{L \left| \frac{1}{{\rm v}_{g1}} - \frac{1}{{\rm v}_{g2}} \right| }.
\end{equation}
This is a general result, which we show below applies to swept-frequency light as well as to a general periodic modulation.

\section{Frequency-swept CW amplifiers}\label{sec:wavelength_sweeping}

Recent work done by White and coauthors \cite{White2012,Petersen2013,Vasilyev2013,White2014} has demonstrated the use of linearly chirped seeding of CW fiber amplifiers in coherent beam combination. Swept frequency amplification was demonstrated to suppress SBS \cite{White2012,Petersen2013,Vasilyev2013}, and this technique was proposed as a method to also increase the mode instability threshold \cite{White2014}.

In this section we analyze STRS suppression for CW signal light that is repetitively swept in wavelength, for example with a triangular or sawtooth waveform. As the wavelength changes, the beat length between the two modes changes due to modal dispersion. If the number of beats changes by one or more, and the change happens rapidly compared to the time required to establish a new temperature distribution, the grating will wash out near the signal output end. Near the signal input end, the grating will be stable; however, near the signal output end the grating periodically moves. If the motion cycles faster than the thermal response time, the grating will be diminished.

We use the criterion that the number of beats must change over the length of the fiber by one half or less. The number of intermodal beats over the length of the fiber is given by
\begin{equation}\label{eqn:num_beats}
N_B = \frac{(\beta_1 - \beta_2)L}{2 \pi},
\end{equation}
where $L$ is the length of the fiber and $\beta_j$ is the propagation constant of mode $j$. Our criterion can be written
\begin{equation}\label{eqn:wavelength_sweep_criterion}
\Delta N_B = \frac{\partial N_B}{\partial \omega}\Delta \omega_{\rm c} = \frac{1}{2}.
\end{equation}
Differentiating Eq.~(\ref{eqn:num_beats}) with respect to $\omega$ gives
\begin{equation}
\frac{\partial N_B}{\partial \omega} = \frac{L}{2 \pi}\left[ \frac{\partial \beta_1}{\partial \omega} - \frac{\partial \beta_2}{\partial \omega} \right].
\end{equation}
The group velocity for mode $j$, ${\rm v}_{gj}$ is defined by
\begin{equation}
\frac{1}{{\rm v}_{gj}} = \frac{\partial \beta_j}{\partial \omega},
\end{equation}
so our criterion from Eq.~(\ref{eqn:wavelength_sweep_criterion}) becomes
\begin{eqnarray}
\Delta \nu_{\rm c} &=& \frac{0.5}{L \left[ \frac{\partial \beta_1}{\partial \omega} - \frac{\partial \beta_2}{\partial \omega} \right]}.
\end{eqnarray}
Or in terms of group velocity, 
\begin{equation}\label{eqn:swept_answer}
\Delta \nu_{\rm c} = \frac{0.5}{L \left[ \frac{1}{{\rm v}_{g1}} - \frac{1}{{\rm v}_{g2}} \right]}.
\end{equation}
Apart from a small difference in the numerical factor, this is identical to Eq.~\ref{eqn:delta_nu_pulses}.

\section{General modulation}\label{sec:general}

We've shown in the previous sections how short pulses and frequency-swept amplifiers give the same criteria for suppression of STRS. In this section, we consider a more general periodic modulation which could be amplitude modulation such as in a series of complex pulses, or it could be periodic phase modulation. Our treatment will be readily extendable to more general cases.

\begin{figure}[htpb]
\centering
\includegraphics[width=0.8\textwidth]{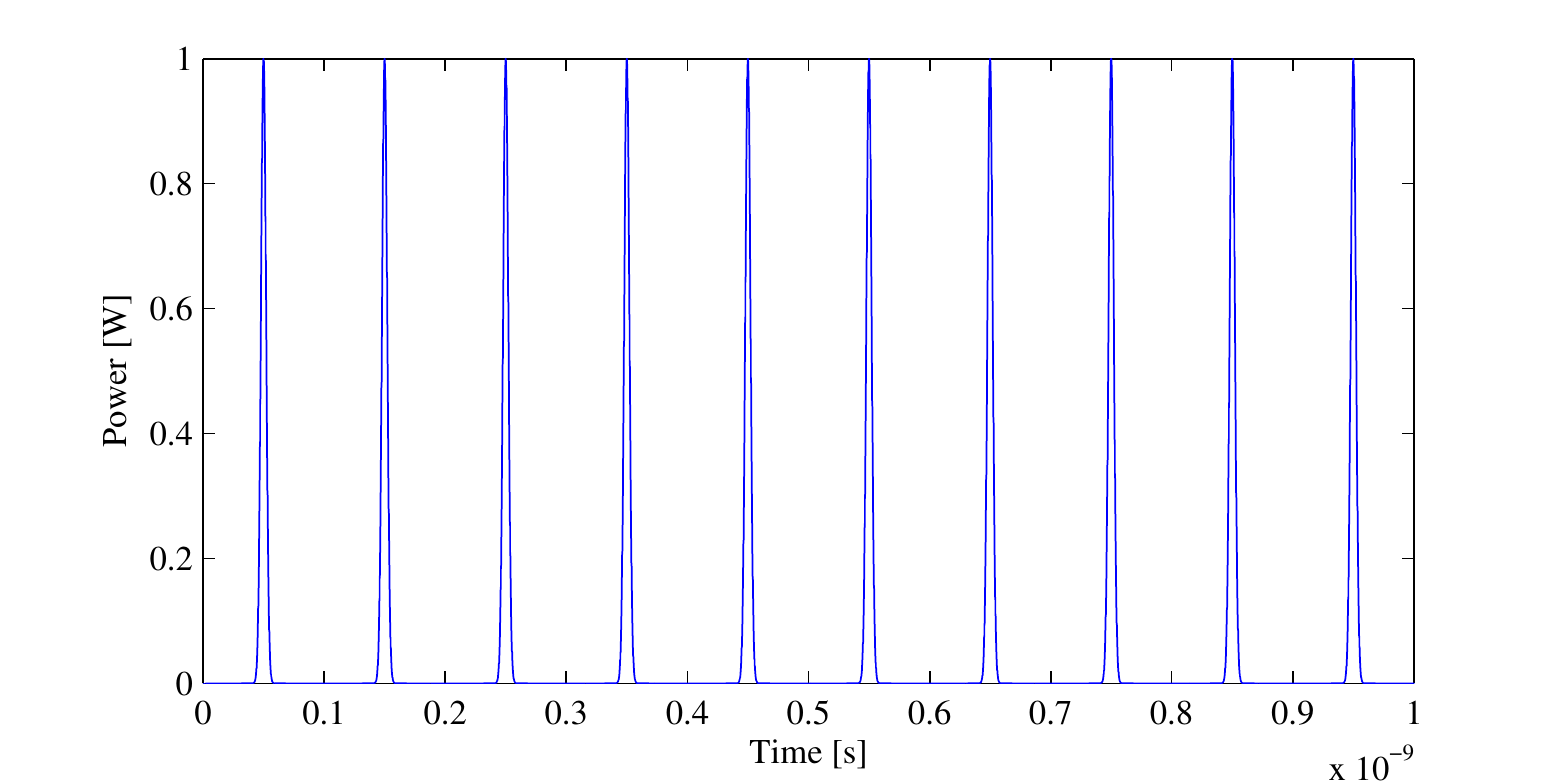}
\caption{\label{fig:pulse_train_timeseries}Time series for a pulse train such as one from a mode-locked laser.}
\end{figure}

\begin{figure}[htpb]
\centering
\includegraphics[width=0.8\textwidth]{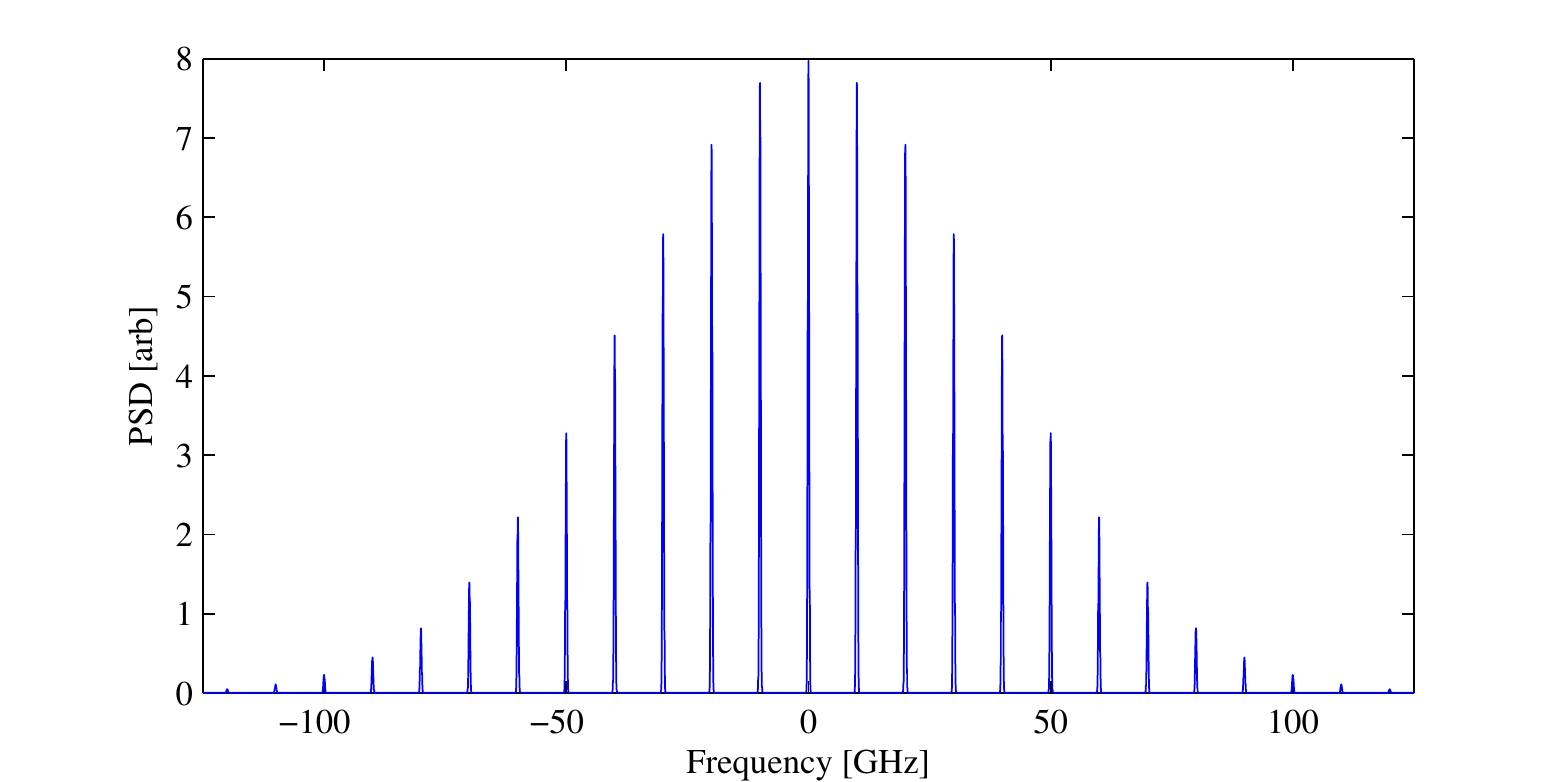}
\caption{\label{fig:pulse_train_spectrum}Spectrum of the pulse train in Fig.~\ref{fig:pulse_train_timeseries}. The Gaussian envelope is set by the Gaussian temporal profile of a single pulse in the train; the comb of frequencies in the envelope come from the separation between pulses. The finite width of each tine in the comb is a by-product of taking the Fourier transform of a finite length of time -- an inifinitely long series of the pulses would fill the Gaussian envelope with a series of true delta functions.}
\end{figure}

To illustrate, consider a train of regularly-spaced, Gaussian pulses similar to one created by a mode-locked laser. The spectrum of such a train of pulses consists of a Gaussian envelope whose width is the inverse of the pulse width filled with a comb of frequencies spaced by the inverse of the inter-pulse spacing. Figure \ref{fig:pulse_train_timeseries} is temporal depiction of such a train of pulses, which have the spectrum depicted in Fig.~\ref{fig:pulse_train_spectrum}.

Suppose that spontaneous thermal Rayleigh scattering (sTRS) seeds LP$_{11}$ with a replica spectra, reduced in amplitude and Stokes shifted to maximize STRS gain. This Stokes shift is much smaller than the frequency spacing of the frequency comb, ensuring that the modulation period is much shorter than the thermal response time.

We can analyze this in frequency space. The temporal profile of LP$_{01}$ can be described by the sum of the appropriately weighted and phased constituent frequency components:
\begin{equation}\label{eqn:E1_1}
E_1(t,z) = \sum_{n=-\infty}^{\infty} E_{1n} e^{-i(\omega_c + n\Delta)t} e^{i \beta_{\rm 1n} z},
\end{equation}
where $\omega_{\rm c}$ is the carrier angular frequency, $\Delta$ is the inverse of the temporal separation between pulses, and $\beta_{\rm 1n}$ is the propagation constant of the $n^{\rm th}$ frequency component given by
\begin{eqnarray}
\beta_{\rm 1n} &=& \beta_{\rm 1c} + \frac{\partial \beta_1}{\partial \omega}n \Delta\nonumber\\
\label{eqn:E1_2}\beta_{\rm 1n} &=& \beta_{{\rm 1c}} + \frac{1}{{\rm v}_{g1}}n\Delta,
\end{eqnarray}
where $\beta_{\rm 1c}$ is the carrier propagation constant for LP$_{11}$.

Combining \autoref{eqn:E1_1} and \autoref{eqn:E1_2} gives,
\begin{equation}
E_{1} (t,z) = \sum_{n=-\infty}^{\infty} E_{1n} e^{-i\big( \omega_c +n\Delta \big) t} e^{i\big( \beta_{1c} + \frac{1}{{\rm v}_{g1}}n \Delta\big) z}.
\end{equation}

Mode 2 has a spectrum identical to mode 1 except it is Stokes-shifted (by $\delta$) and attenuated, so it can be described similarly:
\begin{equation}
E_{2} (t,z) = \sum_{m=-\infty}^{\infty} E_{2m} e^{-i\big( \omega_c +m\Delta-\delta\big) t} e^{i\big( \beta_{2c} + \frac{1}{{\rm v}_{g2}}m \Delta\big) z}.
\end{equation}

Because the temperature grating is caused by quantum defect heating which is related to the irradiance grating, we are interested in the interference between modes 1 and 2, given by $| E_1 + E_2 |^2$. Because $E_1$ and $E_2$ are complex, 
\begin{equation}
| E_1^{} + E_2^{} |^2 = (E_1^{} + E_2^{})(E_1^* + E_2^*) = E_1^{} E^*_1 + E_1^{} E^*_2 + E_1^* E_2^{} + E_2^{} E^*_2,
\end{equation} where the $^*$ indicates complex conjugate. However, because only the cross-terms can produce an irradiance shape that can couple the modes, we can neglect the $E_1^{} E_1^*$ and $E_2^{} E_2^*$ terms keeping only $E_1^{} E_2^* + cc$, which can be written as
\begin{eqnarray}\label{eqn:E1E2_mn}
E_1 E_2^* + cc = \Bigg[ \sum_{n=-\infty}^{\infty} E_{1n} e^{-i\left(\omega_c + n \Delta\right)t} e^{i\left(\beta_{1c} + \frac{1}{{\rm v}_{g1}}n \Delta\right)z} \times \\ \nonumber\sum_{m=-\infty}^{\infty} E_{2m}^* e^{i\left(\omega_c+m\Delta - \delta\right)}e^{-i\left(\beta_{2c} + \frac{1}{{\rm v}_{g2}}m\Delta\right)z}\Bigg] + cc.
\end{eqnarray}
Because frequencies much higher than the inverse thermal diffusion time across the fiber core contribute only weakly to the temperature grating generated by quantum defect heating, we need consider only frequencies near $\omega = \delta$. This dictates that $n=m$, so Eq.~\ref{eqn:E1E2_mn} can be rewritten as
\begin{eqnarray}\label{eqn:E1E2_n}
E_1^{} E_2^* + cc = \Bigg[ \sum_{n=-\infty}^{\infty} E_{1n}^{} e^{-i\left(\omega_c + n \Delta\right)t} e^{i\left(\beta_{1c} + \frac{1}{{\rm v}_{g1}}n \Delta\right)z}\times \\ \nonumber E_{2n}^* e^{i\left(\omega_c+n\Delta - \delta\right)}e^{-i\left(\beta_{2c} + \frac{1}{{\rm v}_{g2}}n\Delta \right)z}\Bigg] + cc,
\end{eqnarray}
\begin{equation}\label{eqn:E1E2_n_product}
E_1^{} E_2^* = \sum_{n=-\infty}^{\infty} E_{1n}^{} E_{2n}^* \exp\Bigg({-i \delta t} + {i \bigg[ (\beta_{1c} - \beta_{2c})} + { n \Delta \bigg( \frac{1}{{\rm v}_{g1}} - \frac{1}{{\rm v}_{g2}} \bigg) \bigg] z} \Bigg).
\end{equation}
For each frequency component $n$, this expression describes an irradiance grating moving toward the output end of the fiber, assuming ($\beta_{1c} - \beta_{2c}$) and $\delta$ have the same sign. Each of the frequency components has a slightly different velocity due to the $n\Delta$ term.

Because we assume the higher order mode is populated by sTRS near the input end of the fiber, $E_{2n}$ is related to $E_{1n}$ simply by
\begin{equation}
E_{2n} = a E_{1n} e^{i\phi}, 
\end{equation}
where $a \ll 1$ and $\phi$ is an arbitrary phase determined by the phase of the sTRS process. Because of this relation, all grating frequency components are in phase at the input end of the fiber. Our criterion for reduced grating strength is that over the length of the fiber, the different gratings develop a phase shift comparable to $\pi$. That is $n$ has a critical value given by
\begin{equation}\label{eqn:general_criterion}
n_c\Delta\Bigg( \frac{1}{{\rm v}_{g1}} - \frac{1}{{\rm v}_{g2}} \Bigg)L = \pi,
\end{equation}
so the critical spectral width (FWHM) is
\begin{equation}
\Delta \nu_c = \frac{n_c \Delta}{2 \pi}.
\end{equation}
Combining this with Eq.~\ref{eqn:general_criterion}, we find the critical linewidth is
\begin{equation}\label{eqn:general_critical_linewidth}
\Delta \nu_c = \frac{0.5}{L\Big(\frac{1}{{\rm v}_{g1}} - \frac{1}{{\rm v}_{g2}}\Big)},
\end{equation}
in agreement with Eq.~\ref{eqn:delta_nu_pulses} and Eq.~\ref{eqn:swept_answer}.

To convert this frequency bandwidth to wavelength bandwidth, we can use the relation that $\Delta \lambda = -\Delta \nu \frac{\lambda^2}{c}$ to find the critical wavelength spread, assuming $\Delta \lambda \ll \lambda$, is
\begin{equation}\label{eqn:pulse_result}
\Delta \lambda_{c} L = \frac{\lambda^2}{c} \frac{0.5}{\left| \frac{1}{{\rm v}_{g1}} - \frac{1}{{\rm v}_{g2}} \right|}.
\end{equation}

Although the example we chose for illustration was a long train of evenly spaced short pulses, the analysis presented applies for any periodic modulation (amplitude or phase). Our assumption that STRS produces a single replica spectrum in LP$_{11}$ can be generalized to assume STRS produces many replica spectra, each with a different frequency shift, but the STRS process amplifies only the replicas in a narrow band centered on the STRS gain maximum. This was illustrated in movies in Ref.~\cite{Smith2014}. In our derivations of Eq.~\ref{eqn:general_critical_linewidth} we did not make any assumption about the relation between $E_{1n}$ and $E_{1m}$ so, although we showed a mode locked train, the derivation applies to arbitrary periodic modulations. Finally, the modulation period can be made as long as desired without changing our conclusion.

\section{Numerical example}

In this section we calculate the critical linewidths, given by Equations \ref{eqn:delta_nu_pulses}, \ref{eqn:swept_answer} and \ref{eqn:pulse_result}, for a few typical step-index fibers. We assume the index step is small, which allows us to use a scalar mode solver but restricts us to linearly-polarized modes such as LP$_{01}$ and LP$_{11}$. Our mode solver uses $\sin-\sin$ expansion. More details of the mode solver can be found in Ref.~\cite{Smith2013a}. In addition to finding the two-dimensional profile of the modes, the mode solver finds propagation constants. Calculating the propagation constants for LP$_{01}$ and LP$_{11}$ for several wavelengths allows us to numerically calculate the group velocities $\nicefrac{\partial \beta}{\partial \omega}$ for each mode as a function of wavelength.

\begin{figure}[htbp]
\centering
\includegraphics[width=0.8\textwidth]{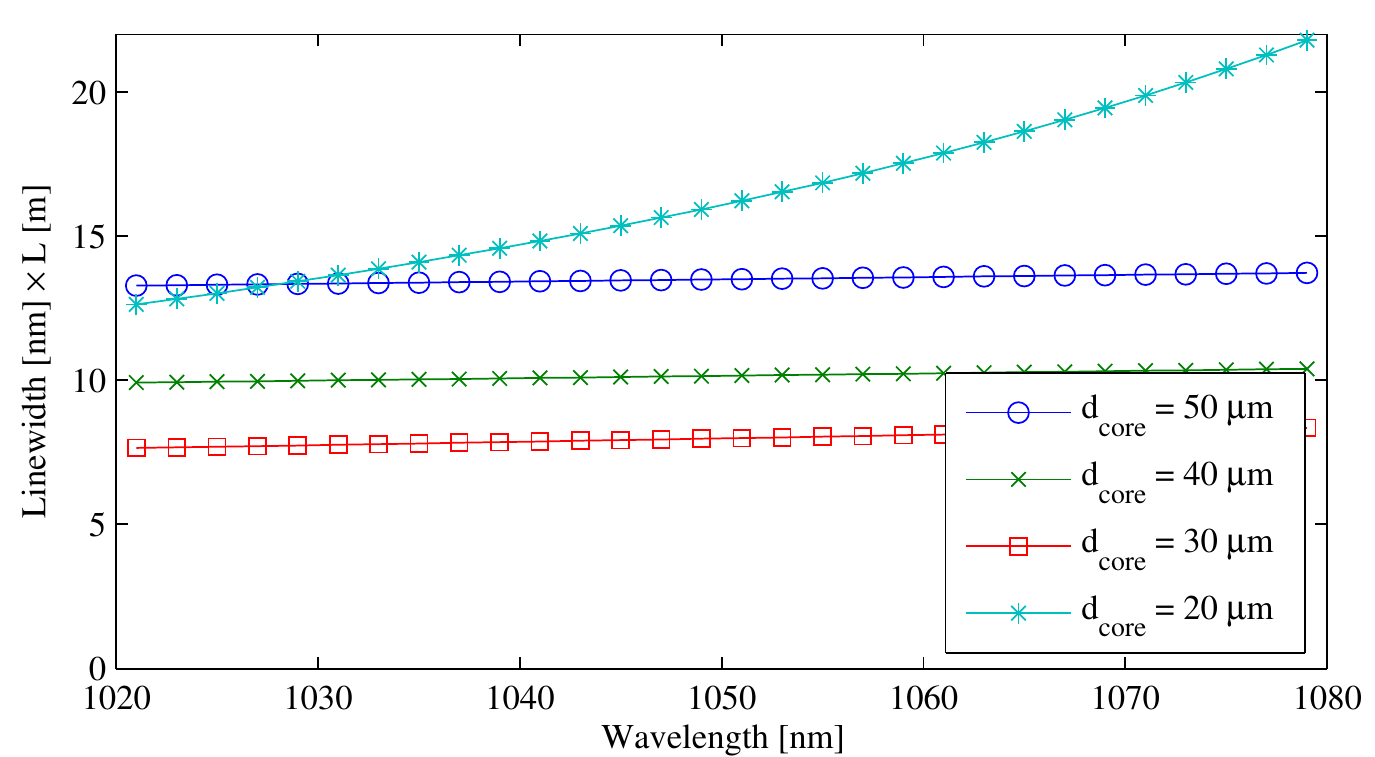}
\caption{\label{fig:linewidth}Product of fiber length and critical linewidth needed to satisfy the criteria described in Sections \ref{sec:pulses}, \ref{sec:wavelength_sweeping} and \ref{sec:general}. Table \ref{table:fibers} lists the index step and numerical aperture for the different core diameters.}
\end{figure}

In Fig.~\ref{fig:linewidth}, the product of linewidth necessary in nanometers and fiber length in meters is plotted as a function of wavelength for several fiber core diameters. We model the fibers listed in Table \ref{table:fibers}. The index step is $0.05$ for all except the 20$\mu$m core diameter because it would otherwise support only one guided mode. For the cladding index we use the Sellmeier equation of silica \cite{Malitson}. The index step is added for the core index.

\begin{table}
\centering
\caption{\label{table:fibers}V-number and N.A. for modeled fibers}
\begin{tabular}{c c c c}
d$_{\rm core}$ & $\Delta n$ & N.A. & V-number \\\hline
20 $\mu$m & 0.03 & 0.09 & 5.5\\
30 $\mu$m & 0.01 & 0.05 & 4.8\\
40 $\mu$m & 0.01 & 0.05 & 6.4\\
50 $\mu$m & 0.01 & 0.05 & 7.8\\
\end{tabular}
\end{table}

\section{Discussion}

While this paper is intended to give an estimate of the gain reduction, it is difficult to give an exact answer for a number of reasons. For instance, the distribution of STRS gain along the fiber amplifier must be taken into account. In a co-pumped amplifier, most of the STRS gain occurs in the first half of the fiber -- meaning if we start washing out the grating near the signal output end, the bulk of the gain from STRS will be nearly unchanged. To complicate matters, constructing and running a numerical model which rigorously treats the matters discussed above is prohibitively computationally expensive. A rigorous model which must consider bandwidths on the order of terahertz will require a time resolution much finer than that required for the STRS process, which happens at a few hundred hertz, meaning billions of time points need to be included. A coupled mode model could include linewidths and group velocity.

Finally, we've used modes LP$_{01}$ and LP$_{11}$ because we've found the largest STRS gain between those two modes in step-index fibers \cite{Smith2012,Hansen2013}, but obviously other mode pairs can be considered.

\end{document}